\documentclass[twocolumn,showpacs,prl]{revtex4}

\usepackage{graphicx}

\begin{document}

\def\newline{\ \par}
\def\br{{\bf r}}
\def\bq{{\bf q}}
\def\bx{{\bf x}}
\def\by{{\bf y}}
\def\bA{{\bf A}}
\def\bS{{\bf S}}
\def\etal{{\em et al. }}

\title{Different ocular dominance map formation influenced by \\
  orientation preference columns in visual cortices}
\author{Myoung Won Cho}
\email{mwcho@postech.edu}
\author{Seunghwan Kim}
\email{swan@postech.edu}
\affiliation{
  Asia Pacific Center for Theoretical Physics and Nonlinear
  $\&$ Complex Systems Laboratory - NRL, \\
  Department of Physics, Pohang University of Science and
  Technology, Pohang, Gyeongbuk, 790-784, Korea
}
\begin{abstract}
In animal experiments, the observed orientation preference (OP) and ocular
dominance (OD) columns in the visual cortex of the brain show various pattern
types.
Here, we show that the different visual map formations in various species are
due to the crossover behavior in anisotropic systems composed of
orientational and scalar components such as easy-plane Heisenberg models.
We predict the transition boundary between different pattern types with the
anisotropy as a main bifurcation parameter, which is consistent with
experimental observations.
\end{abstract}
\maketitle

The highly ordered structure in the mammalian visual cortex has attracted much
attention from theoretical neurobiologists and has been thoroughly studied with
the expectation of providing the basis for neural dynamics and computational
models.
Though most models of the visual map formation are based on common postulates,
such as Hebbian synapses, connections or competitions between neighboring
neurons and synaptic normalization, there are quite a number of
successful models with unique mechanisms~\cite{Erwin1995,Swindale1996}.
The Hamiltonian models with spin variables were proposed for the visual map
formation with a striking analogy with the physical systems, such as
magnetism~\cite{Tanaka1989,Cowan1991}.
Recently, the characteristics of visual maps are systematically explored
through the statistical properties of vortices in magnetism~\cite{Cho2004A}.
The spin-like Hamiltonian models represent essential ingredients of neural
interactions in the visual map formation without paying much attention to
particular neural control mechanisms and can be shown to exhibit common
statistical properties of the vortex formation as in other development
models~\cite{Cho2004C}.

As vast experimental data on visual maps are accumulated, the various patterns
in visual cortices of different animals have drawn much interest with the
expectation of testing different neural models experimentally~\cite{LeVay1985,
Obermayer1993,Lowel1987a,Anderson1988,Crowley1999,Lowel1987b,Bosking1997}.
The observed visual patterns can be classified as at least three different
types.
In macaque monkeys, the OD columns form parallel bands of regular spacing with
relatively few branching points that are mainly oriented perpendicular to area
boundaries~\cite{LeVay1985}.
The degree of OD segregation is strong and the typical average spacing of OD,
$\Lambda_{OD}$, is larger than that of the OP columns, $\Lambda_{OP}$
($\Lambda_{OD}>\Lambda_{OP}$)~\cite{Obermayer1993}.
In cats and ferrets, OD columns form an array of beaded bands exhibiting only a
weak tendency of elongation perpendicular to area
boundaries~\cite{Lowel1987a,Anderson1988,Crowley1999}.
The degree of OD segregation is intermediate with $\Lambda_{OD}<\Lambda_{OP}$
~\cite{Lowel1987b}.
In the case of tree shrews, the OD segregation is very weak or absent, while
more stripe-like patterns are observed in extensive OP columnar regions with
low densities of orientation centers~\cite{Bosking1997}.

Some experimental works show that the OP and OD patterns are structurally
correlated.
The singular points, so-called pinwheels, in OP columns tend to align with the
centers of OD bands and the iso-orientation contours intersect the borders of
OD bands at right or steep angles~\cite{Obermayer1993}.
Such correlations between two columns occur due to the normalization of the
synapse strength~\cite{Grossberg1994}.
The average response of the orientation selectivity vanishes at singularities,
so that the strong OD components can develop near pinwheels.
The orthogonality between contours of the columnar patterns can be explained
intuitively through the isotropic Heisenberg model~\cite{Cho2004A}.
The different visual pattern types originate from both the synaptic
normalization and the anisotropic interaction between two columns.
The parameter $\lambda$, a measure of the anisotropy between the columns, turns
out to be crucial in determining not only the pinwheel stability but also OD
pattern types and OP pattern regularity.
We estimate the anisotropy for each species from the typical spacing of OP and
OD patterns from experiments.
We calculate the threshold of anisotropy between different OD pattern types and
construct the phase diagram, which are consistent with experimental
observations.

One of reasons why so many different neural models lead to the successful
formation of visual maps is that the typical characteristics of self-organizing
maps are determined by the topology of lattice and feature space rather than
the detailed cortical modification rules.
The statistical properties of emergent cortical maps can be predicted using
only minimal mathematical constraints such as the symmetry.
In the fiber bundle map (FBM) representation method, the phases of feature
components are described by the continuous group corresponding to the manifold
in the feature space~\cite{Cho2004A,Cho2004C}.
Because OP columns have $O(2)$ (or $U(1)$) symmetry, the energy functions for
the OP map formation should be invariant under gauge transformations and take a
general form
\begin{eqnarray} \label{eq:OP}
  E_{OP}=\int d\br\left\{\frac{v}{2}|(\nabla-i\bA)\psi_{OP}|^2
    -\frac{m^2}{2}|\psi_{OP}|^2+\cdots\right\}
\end{eqnarray}
for the orientational feature components
$\psi_{OP}(\br)=(q(\br)\cos2\phi(\br),q(\br)\sin2\phi(\br))$ (or
$q(\br)e^{2i\phi(\br)}$) with the preferred angle $\phi(\br)$ and the degree of
preference $q(\br)$ at the cortical location $\br$.
This can be obtained also from other map formation models after a continuum
approximation and can explain most typical properties of OP patterns found in
experiments and simulations~\cite{Cho2004A}.

Biologically the cortical modules are composed of several layers and the
different feature components are found from the selective response properties
of corresponding layers.
Each layer can have a different synaptic strength, so that the energy for the
combined OP and OD map formation takes an anisotropic form
\begin{eqnarray} \label{eq:OP_OD}
  E&=&\int d\br \left\{ \frac{v_{OP}}{2}|(\nabla-i\bA_{OP})\psi_{OP}|^2
    -\frac{m_{OP}^2}{2}|\psi_{OP}|^2 \right. \ \ \ \ \\
  && \ \ \ \ \left. +\frac{v_{OD}}{2}|(\nabla-i\bA_{OD})\psi_{OD}|^2
    -\frac{m_{OD}^2}{2}|\psi_{OD}|^2\right\} \nonumber
\end{eqnarray}
including the scalar component representation of OD $\psi_{OD}$.
The restriction imposed by the normalization for the total feature vector
$\psi(\br)=(\psi_{OP}(\br),\psi_{OD}(\br))$ is that
\begin{eqnarray}
  |\psi(\br)|^2=|\psi_{OP}(\br)|^2+|\psi_{OD}(\br)|^2=const
\end{eqnarray}
for all $\br$.
The normalization and the equilibrium ($\nabla^2\psi_\mu\sim 0$ for $\mu=OP$
and $OD$) conditions lead to the orthogonality between OP and OD contour lines,
that is $\nabla \psi_{OP}\cdot\nabla \psi_{OD}\sim 0$.
Now, we define a parameter $\lambda=m_{OD}^2/m_{OP}^2$ describing the
anisotropy between two columns.
Then Eq.(\ref{eq:OP_OD}) has two different solutions depending on the
anisotropy $\lambda$.
If $\lambda<1$ (or $\lambda>1$), it has a ground state at the energy density
$-m_{OP}^2/2$ (or $-m_{OD}^2/2$) with weak OD segregations
$\langle\psi_{OD}^2\rangle=0$ (or strong OD segregations
$\langle\psi_{OD}^2\rangle=|\psi|^2$).
The previous studies in magnetism have shown that there exists another
threshold $\lambda_c$ when features are composed of orientational and scalar
components.
The statistical properties of visual maps are closely related to those of the
classical two-dimensional anisotropic Heisenberg model (CTDAHM), described by
the Hamiltonian
\begin{eqnarray}
  H=-J\sum_{\langle ij\rangle}(S_i^xS_j^x+S_i^yS_j^y+\lambda S_i^zS_j^z),
\end{eqnarray}
for $J>0$.
The possibility of two vortex types, named in-plane and out-of-plane, in the
CTDAHM was first discussed by Takeno and Homma~\cite{Takeno1980}.
Different values of the critical anisotropy $\lambda_c$, above which
out-of-plane vortices become stable, were found to exist and depend on the
lattice type ($\lambda_c\approx 0.72$, $0.82$, $0.62$ for square, honeycomb,
and triangular lattices, respectively)~\cite{Gouvea1989}.

\begin{figure}[t]
\begin{minipage}[b]{7.7cm} \begin{center}
\begin{minipage}[b]{3.8cm} \begin{center}
  \includegraphics[width=3.8cm]{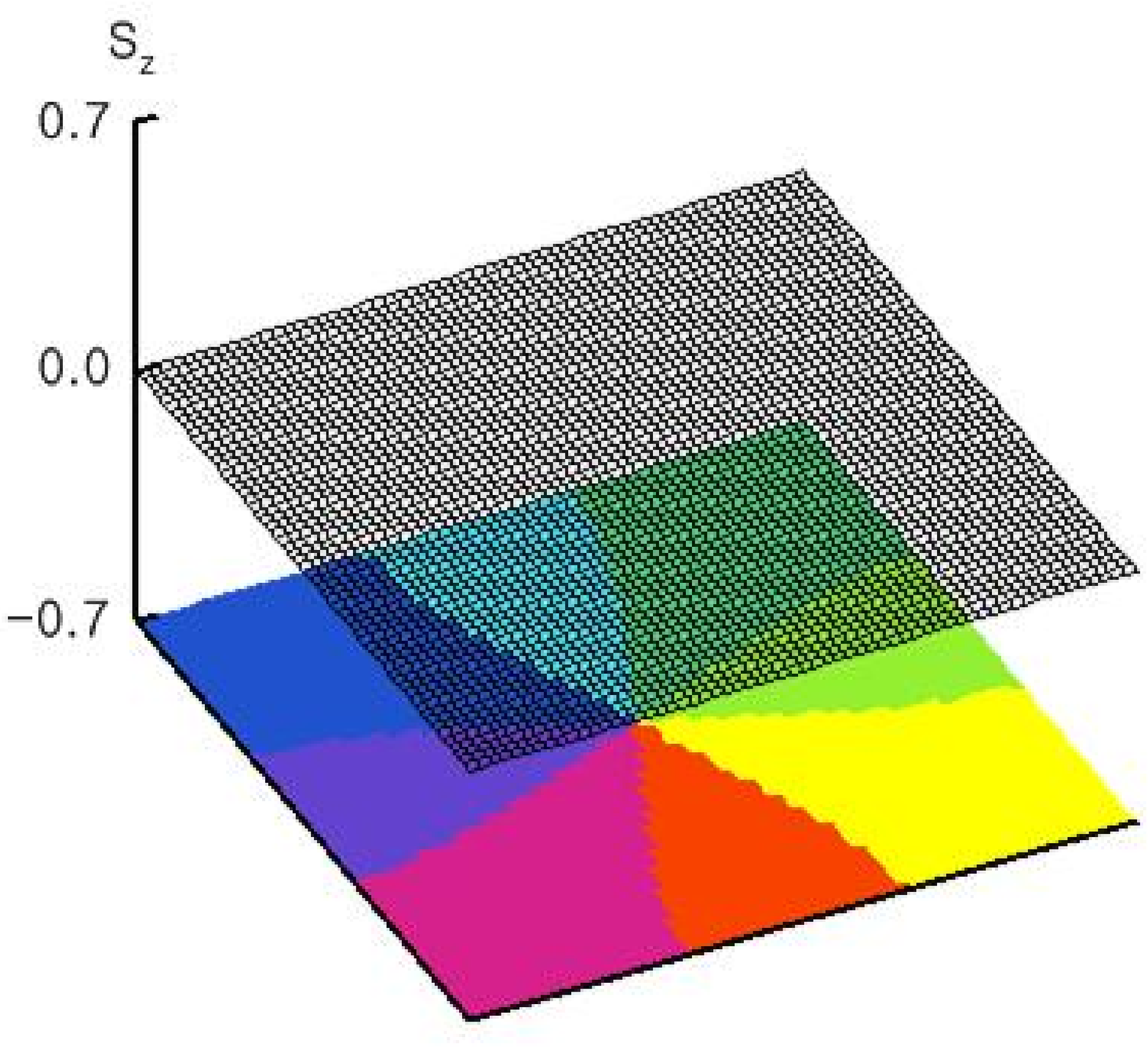} (a) $k=0.2$, $\lambda=0.63$
\end{center} \end{minipage}
\begin{minipage}[b]{3.8cm} \begin{center}
  \includegraphics[width=3.8cm]{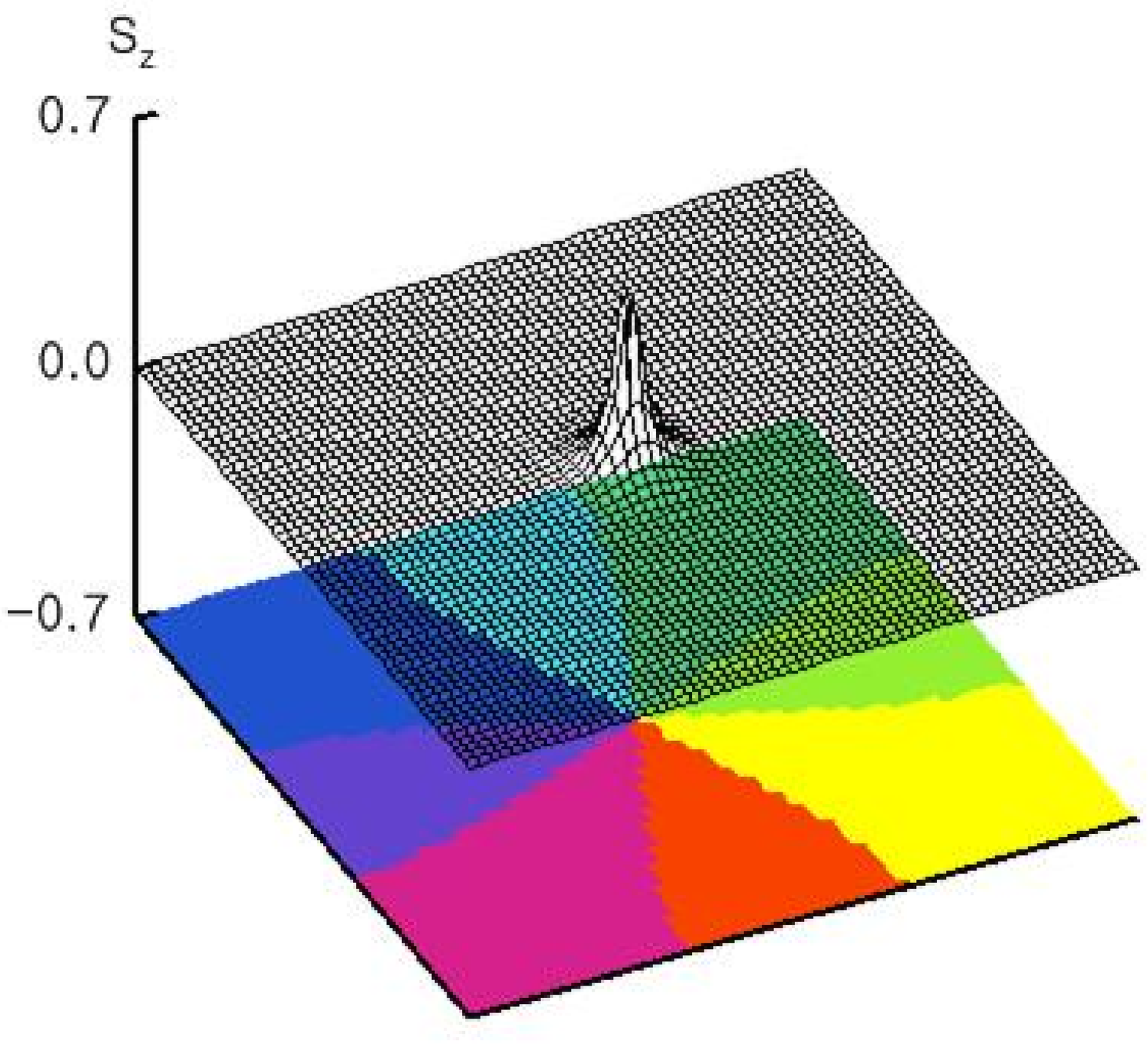} (b) $k=0.2$, $\lambda=0.67$
\end{center} \end{minipage}
\begin{minipage}[b]{3.8cm} \begin{center}
  \includegraphics[width=3.8cm]{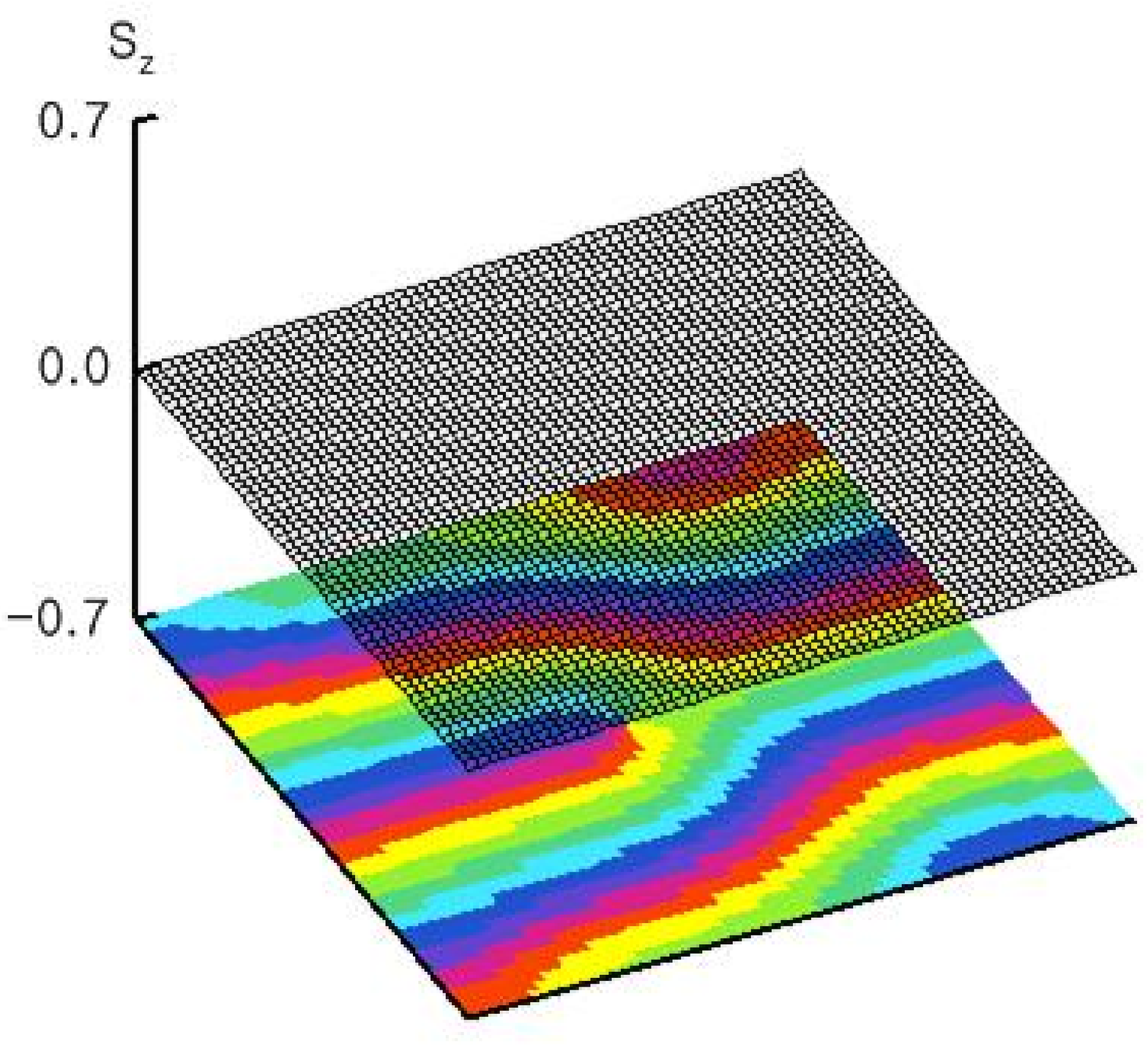} (c) $k=0.3$, $\lambda=0.63$
\end{center} \end{minipage}
\begin{minipage}[b]{3.8cm} \begin{center}
  \includegraphics[width=3.8cm]{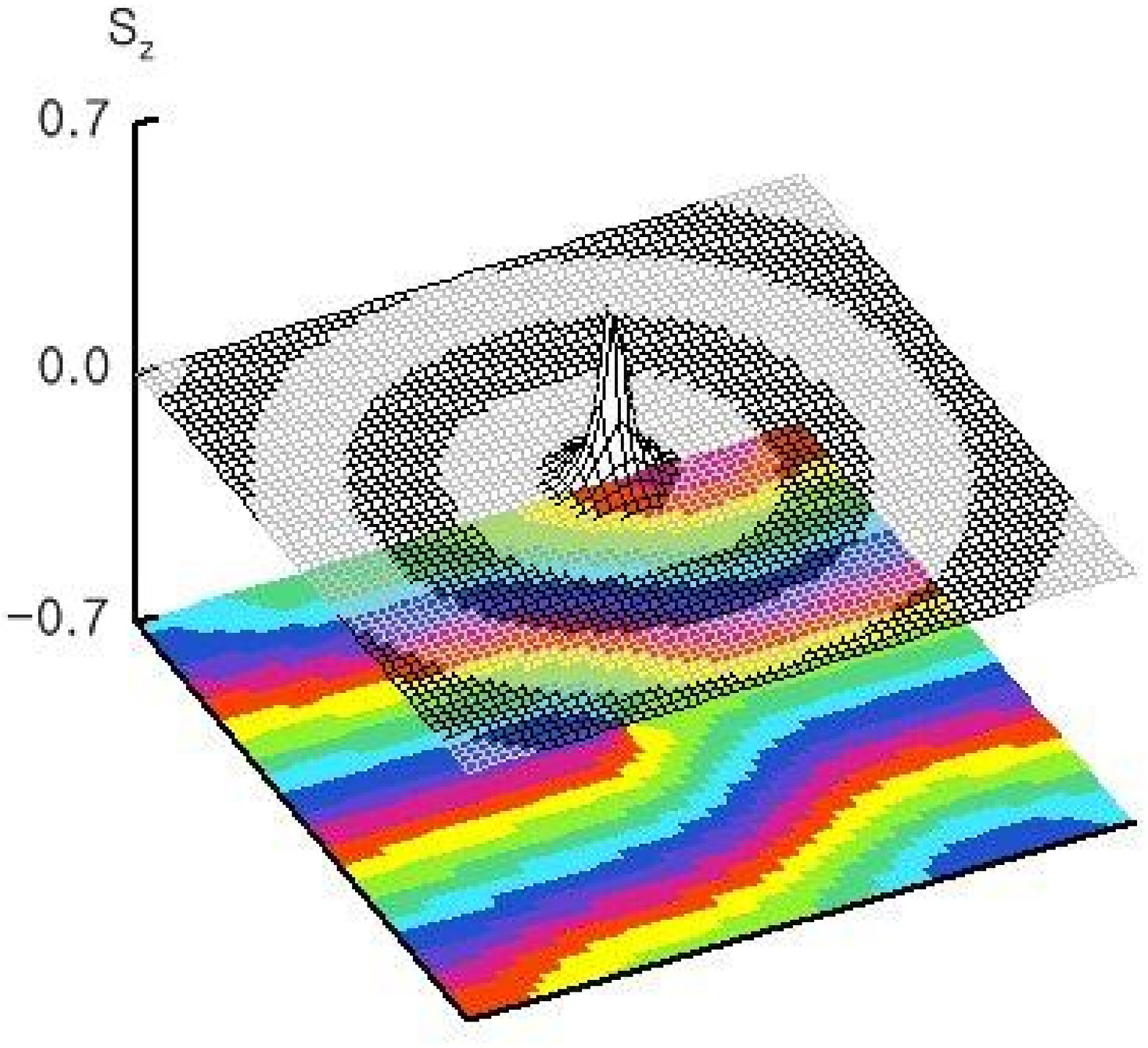} (d) $k=0.3$, $\lambda=0.67$
\end{center} \end{minipage}
\end{center} \end{minipage}
\ 
\begin{minipage}[b]{0.5cm}
  \includegraphics[width=0.5cm]{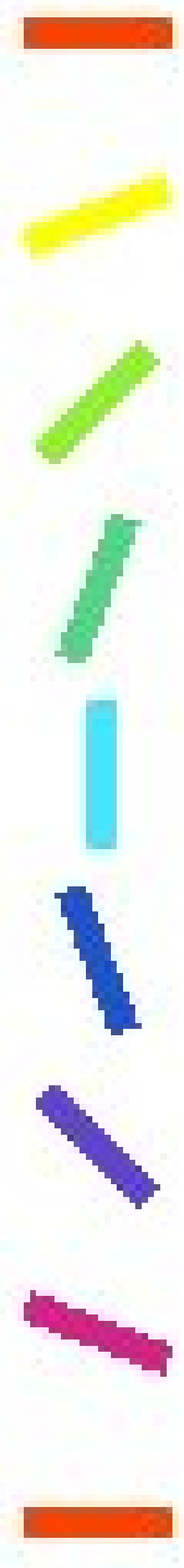}
  \vspace{2.2cm}
\end{minipage}
\caption{ \label{fig:vortices}
  Development of the scalar peak near the orientational singularity.
  The preferred angle $\phi$ is denoted in 8 colors and the ocular dominance
  $S_z$ is denoted in black and white wired frames, where the black denotes the
  region with $S_z\ge 0$ and the white, otherwise.
  Simulations are done for the Hamiltonian in Eq.(\ref{eq:Hamiltonian}) with
  $\sigma_{OD}^2=\sigma_{OP}^2=6$, $k_{OD}=k_{OP}$
  ($\lambda=\varepsilon_{OD}/\varepsilon_{OP}$).
  ($60\times 60$ lattice and a non-periodic boundary condition)
}
\end{figure}


In this Letter, the anisotropic Heisenberg model with distance-dependent
interactions is used for simulations :
\begin{eqnarray} \label{eq:Hamiltonian}
  H=-\sum_{i,j}\left\{D_{OP}(r_{ij})(S_i^xS_j^x+S_i^yS_j^y)
    +D_{OD}(r_{ij})S_i^zS_j^z\right\}
\end{eqnarray}
for the preferred angle $\phi_i=(1/2)\tan^{-1}(S_i^y/S_i^x)$, the OD $S_i^z$
with normalization to a unit modulus ($|\bS_i|^2=1$).
This satisfies also the energy form in Eq.(\ref{eq:OP_OD}) in a continuum
approximation.
Other development models, such as the elastic-net model or Kohonen's SOFM
algorithm , can be rewritten into the spin-like Hamiltonian
model~\cite{Cho2004C}.
Usually $D(\bx,\by)=D(|\bx-\by|)$ is positive in short-range and negative in
long-range, that is, of the Mexican hat type.
The typical parameters in Eq.(\ref{eq:OP}) or (\ref{eq:OP_OD}) and the
exact anisotropy $\lambda$ are determined by the actual form of $D(r)$.
In general, the interaction strength within each column is proportional to the
activity strength $\varepsilon_\mu$, so that
$\lambda\propto\varepsilon_{OD}/\varepsilon_{OP}$,
and to the number of interacting pairs, so that
$\lambda\propto\sigma_{OD}^2/\sigma_{OP}^2$ in two-dimension for the
cooperation range $\sigma_\mu$.
We take the form of the interaction function
\begin{eqnarray}
  D_\mu(r)=\varepsilon_\mu\left(1-k_\mu\frac{r^2}{\sigma_\mu^2}\right)
    \exp\left(-r^2/2\sigma_\mu^2\right)
\end{eqnarray}
for an inhibitory strength $k_\mu$.
Single vortex simulations in Fig.~\ref{fig:vortices} show the structure of two
vortex types.
When $\lambda$ is above the threshold $\lambda_c$, a peak in the OD map starts
to develop near a singular point in the OP map (Fig.~\ref{fig:vortices}{\em b}
and \ref{fig:vortices}{\em d}).
When there exists a weak inhibitory activity, we have ferromagnetic solutions
with both a scalar peak and and an orientational singularity
(Fig.~\ref{fig:vortices}{\em b}), which are the typical features of in-plane
and out-of-plane vortices observed in the CTDAHM.
Similar results can be obtained in different map formations models, if they are
modified to have an anisotropy between columns.

\begin{figure}
  \includegraphics[width=8.5cm]{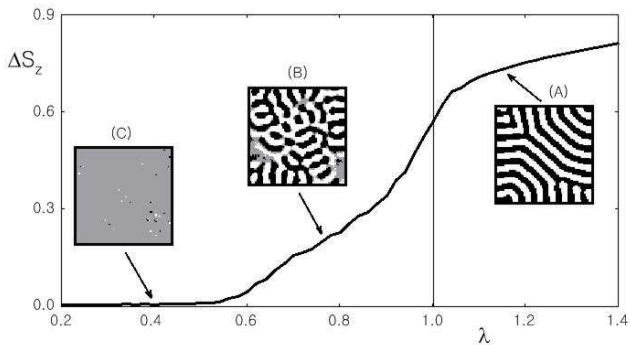}
\caption{ \label{fig:segregation}
  (a) Different OD pattern formations depending on the anisotropy $\lambda$.
  The gray color in boxes denotes the weak segregation region ($|S_z|\le 0.01$).
  Parameters are $\sigma_{OP}^2=\sigma_{OD}^2=6$, $k_{OP}=k_{OD}=1.0$
  ($\lambda=\varepsilon_{OD}/\varepsilon_{OP}$).
  Maps are generated using the Metropolis algorithm at zero temperature, with a
  non-periodic boundary condition and an initially random state in the
  $70\times 70$ lattice.
}
\end{figure}

Fig.~\ref{fig:segregation} shows the emergence of three different
pattern types in OD columns; Type (A) for $\lambda>1$, parallel bands with
strong OD segregation, Type (B) for $\lambda_c<\lambda<1$, beaded bands with
intermediate OD segregation, and Type (C) for $\lambda<\lambda_c$, the absence
of OD segregation.
Types (A), (B), and (C) in Fig.~\ref{fig:segregation} correspond to observed OD
patterns in macaque monkeys, cats, and tree shrews, respectively.
We measure the degree of OD segregation in terms of the standard deviation of
$S_z$ at a given time that is averaged over 10 trial evolutions for a given
$\lambda$.
We find that the OD segregation strength increases as $\lambda$ increases.
The in-plane vortices are unstable under small thermal fluctuations for
$\lambda<1$.
The anisotropy $\lambda$ is related to the pinwheel stability and connected to
the measured pinwheel density for different animals~\cite{Cho2004A,Wolf1998}.
Beaded bands are not permanent structure as well and disappear as pinwheels
annihilate during the map development process.


\begin{figure}[b]
\begin{minipage}[b]{4.0cm}
\includegraphics[width=4.0cm]{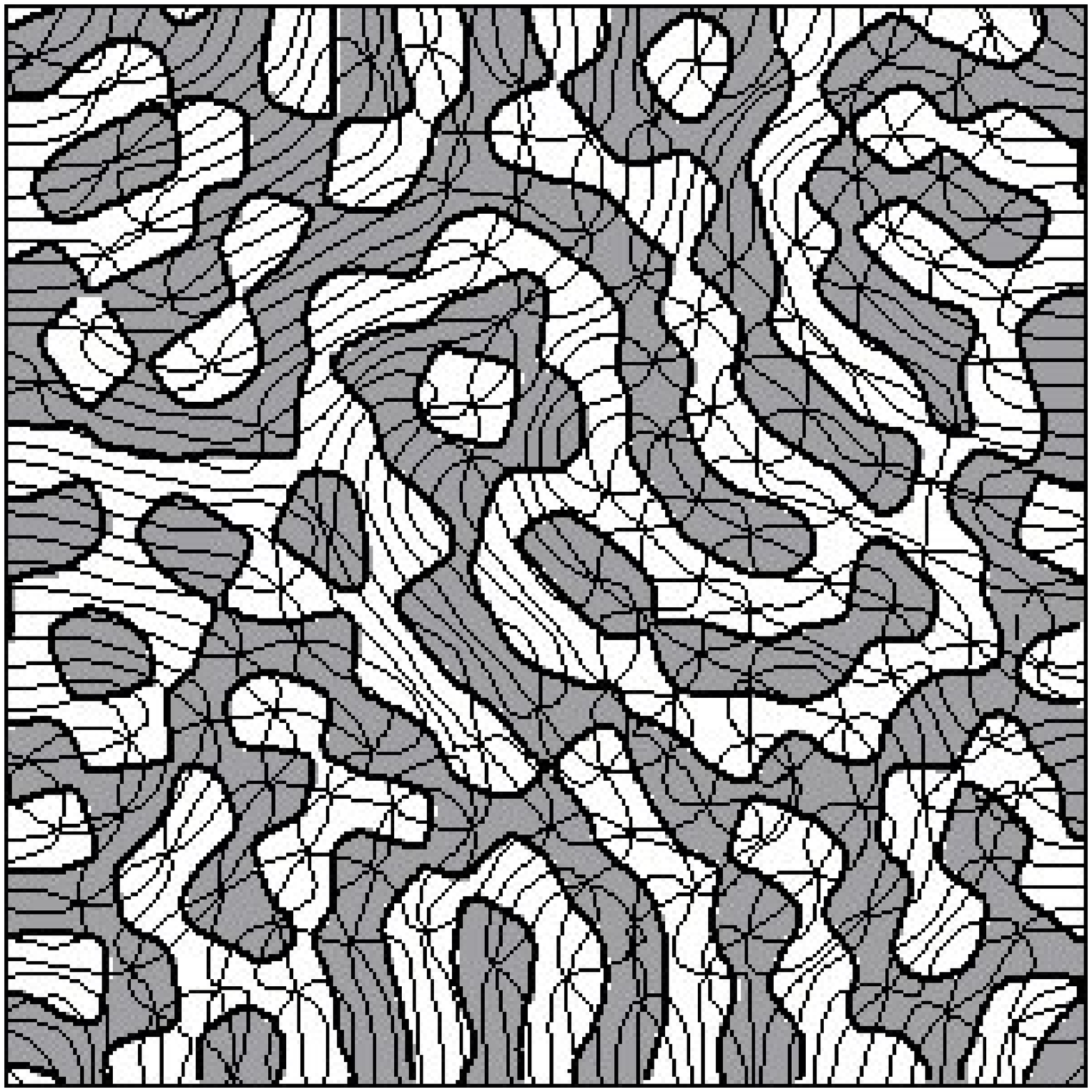} (a)
\end{minipage}
\ 
\begin{minipage}[b]{4.0cm}
\includegraphics[width=4.0cm]{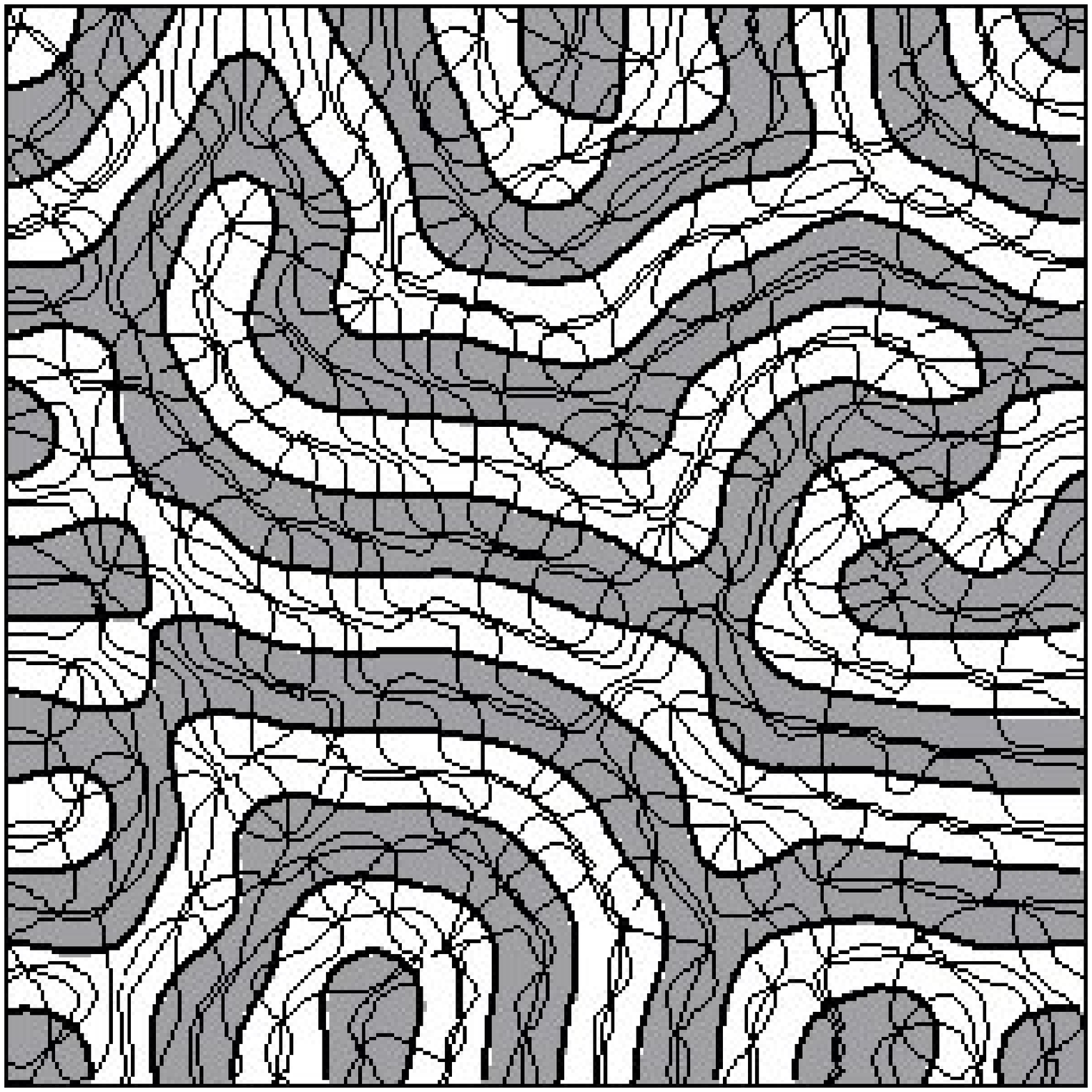} (b)
\end{minipage}
\caption{ \label{fig:simulations}
  Simulations of OD and OP maps for
  (a) $\Lambda_{OD}<\Lambda_{OD}$ ($\sigma_{OD}^2=6.7$, $\sigma_{OP}^2=10.0$)
  and
  (b) $\Lambda_{OD}>\Lambda_{OP}$ ($\sigma_{OD}^2=13.8$, $\sigma_{OP}^2=10.0$).
  Parameters are $\varepsilon_{OD}=\varepsilon_{OP}$ and $k_{OP}=k_{OD}=0.7$,
  so $\lambda=\sigma_{OD}^2/\sigma_{OP}^2=\Lambda_{OD}^2/\Lambda_{OP}^2$
  ($100\times 100$ lattice with a non-periodic boundary condition).
}
\end{figure}


For $\lambda>1$, which corresponds to the Ising model region, the OD columns
play a dominant role in the overall map formations.
Parallel bands appear in a whole range of the OD map and make a right or
steep angle with area boundaries, which are the typical properties of OD maps
generated without OP columns.
In the OP columns, singular points and periodic patterns still exist, but are
much less regular.
The distribution of iso-orientation contours is not uniform with locally dense
or sparse regions (see Fig.~\ref{fig:simulations}{\em b}).
The rotational plane defined from the gradient directions in the feature space
helps to understand this point.
For $\lambda=1$, the map formation process is balanced between two stationary
solutions with the normal vector of the rotational plane slanted at
$|\hat{n}_x|=|\hat{n}_y|=|\hat{n}_z|=1/\sqrt{3}$,
so that $\Delta S_z=\sqrt{1-|\hat{n}_z|^2}/\sqrt{2}=1/\sqrt{3}\simeq0.577$,
as seen in Fig.~\ref{fig:segregation}.
For $\lambda>1$ (or $\lambda<1$), the rotational plane lies more vertically (or
horizontally), so that $\Delta S_z$ approaches $1$ (or $0$) as the map
formation progresses.
If the rotational plane becomes too slanted, the circular trajectories are
projected to elongated ellipses in the $S_x-S_y$ plane, leading to
irregularities in $\phi$ coordinates.

In the real brain, it is not certain which factor among various parameters such
as the activity rate $\varepsilon_\mu$, the cooperation range $\sigma_\mu$ and
the inhibitory strength $k_\mu$, is a major contributor to the anisotropy
between OP and OD columns.
We can obtain some insights on this by comparing the typical spacings
$\Lambda_{OP}$ and $\Lambda_{OD}$ measured in animal experiments.
Various models predict that the typical spacing $\Lambda_\mu$ increases in
proportion to the cooperation range
$\sigma_\mu$ ($\Lambda_\mu\propto\sigma_\mu$) and decreases for stronger
inhibitory strength $k_\mu$~\cite{Obermayer1992,Wolf2000,Cho2004A}.
If OP and OD columns are different only in the activity rate $\varepsilon_\mu$,
the typical spacings are the same, that is $\Lambda_{OD}=\Lambda_{OP}$ for all
$\lambda$.
If they differ only in $k_\mu$, the typical spacing of OD bands is usually
smaller than that of OP slabs when the strong OD segregation occurs
($\Lambda_{OD}<\Lambda_{OP}$ for $\lambda>1$).
However, if they differ only in $\sigma_\mu$, the typical spacing of OD bands
is larger than that of OP slabs when strong OD segregation occurs
($\Lambda_{OD}>\Lambda_{OP}$ for $\lambda>1$).
Among these possibilities, we find that the last case is consistent with the
animal data in cats~\cite{Lowel1987a,Lowel1994,Lowel1987b,Albus1979,Diao1990},
ferrets~\cite{Crowley1999,Rao1997} and macaque monkeys~\cite{Obermayer1993}.
This leads us to use the cooperation ranges between columns as the major
bifurcation parameter for investigating different visual pattern types in
animals.
Fig.~\ref{fig:simulations} shows the results of simulations using the typical
spacings $\Lambda_{OP}$ and $\Lambda_{OD}$ measured in animal experiments.
For $\Lambda_{OD}<\Lambda_{OP}$, more beaded bands in OD columns emerge near
the pinwheel centers as in the visual maps in cats
(Fig.~\ref{fig:simulations}{\em a}), whereas for $\Lambda_{OD}>\Lambda_{OP}$,
parallel bands in OD columns emerge in a whole region as in macaque monkeys
(Fig.~\ref{fig:simulations}{\em b}).

\begin{figure}[t]
\includegraphics[width=7.5cm]{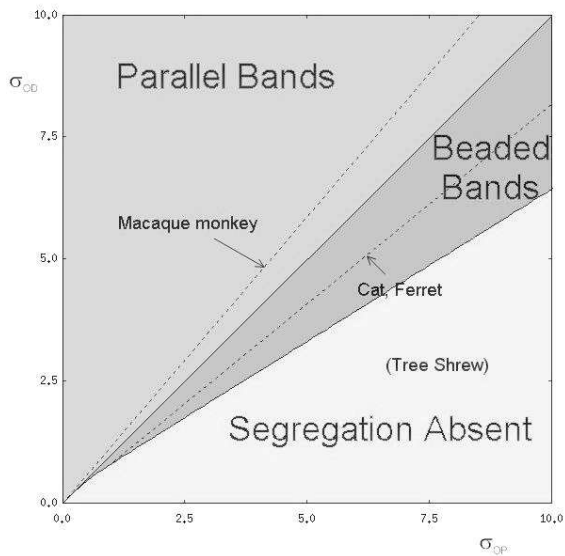}
\caption{ \label{fig:phase}
  The phase diagram of OD patterns according to the cooperation ranges,
  $\sigma_{OD}$ and $\sigma_{OP}$, where dotted lines denote the experimental
  data for various animals.
}
\end{figure}

The exact calculation of the critical anisotropy $\lambda_c$, above which the
out-of-plane vortices appear, is also an important problem in
magnetism~\cite{Costa1996}.
Unfortunately, any continuum theory fails in calculation of $\lambda_c$ because
of the singularity near the vortex core~\cite{Wysin1994}.
The development of out-of-plane components depends on not only the exact form
of the interaction function $D(r)$ but also other factors such as lattice
types, vortex distributions, etc.
The boundary between the ``OD segregation absent'' and ``Beaded bands'' regions
cannot be distinguished sharply in both experiments and simulations because the
development of OD components is weak near this boundary.
The analytic and simulational methods in the CTDAHM can be extended to
calculate the critical anisotropy precisely for various cases~\cite{Cho2004B}.
We find that the stability boundary of the out-of-plane vortex takes the
form $\sigma_{OD}\simeq\alpha\sigma_{OP}+\beta$.
Since the computed value of $\beta$ is small, the region for the beaded band
pattern can be described by $\alpha<\sigma_{OD}/\sigma_{OP}<1$, where
$\alpha\sim 0.6$ for $\lambda=\sigma_{OD}^2/\sigma_{OP}^2$.
This agrees well with the experimental data, where
$\Lambda_{OD}/\Lambda_{OP}\sim0.82$ for cats and ferrets.
Fig.~\ref{fig:phase} shows the phase diagram of different OD pattern formations
in the parameter of cooperation ranges, $\sigma_{OP}$ and $\sigma_{OD}$.
The experimentally derived curves for macaque monkeys, cats, and ferrets lie
well within regions of our predictions.

In this Letter, we showed the influence of the correlation between OP and OD
columns on the overall visual map formation, and explained the observed
experimental data successfully.
Moreover, we also showed that the typical characteristics of emergent cortical maps
can be described well by the solutions from the general field theory with
minimal constraints.
Our study of the visual map formation suggests another possibility in modeling
physical neuronal dynamics with physical models.
In spite of the complex circuitry and nonlinear dynamics in neural systems, the
neural behavior at high levels may obey simple and general rules, which can be
related to the noble theory of statistical physics.
We expect to find more applications of physics-based methodology in
interpreting diverse neural phenomena and constructing noble computational
architecture.


This work was supported by the Ministry of Science and Technology and the
Ministry of Education.

\bibliography{od}

\begin{thebibliography}{26}
\expandafter\ifx\csname natexlab\endcsname\relax\def\natexlab#1{#1}\fi
\expandafter\ifx\csname bibnamefont\endcsname\relax
  \def\bibnamefont#1{#1}\fi
\expandafter\ifx\csname bibfnamefont\endcsname\relax
  \def\bibfnamefont#1{#1}\fi
\expandafter\ifx\csname citenamefont\endcsname\relax
  \def\citenamefont#1{#1}\fi
\expandafter\ifx\csname url\endcsname\relax
  \def\url#1{\texttt{#1}}\fi
\expandafter\ifx\csname urlprefix\endcsname\relax\def\urlprefix{URL }\fi
\providecommand{\bibinfo}[2]{#2}
\providecommand{\eprint}[2][]{\url{#2}}

\bibitem[{\citenamefont{Erwin et~al.}(1995)\citenamefont{Erwin, Obermayer, and
  Schulten}}]{Erwin1995}
\bibinfo{author}{\bibfnamefont{E.}~\bibnamefont{Erwin}},
  \bibinfo{author}{\bibfnamefont{K.}~\bibnamefont{Obermayer}},
  \bibnamefont{and} \bibinfo{author}{\bibfnamefont{K.}~\bibnamefont{Schulten}},
  \bibinfo{journal}{Neural comput.} \textbf{\bibinfo{volume}{7}},
  \bibinfo{pages}{425} (\bibinfo{year}{1995}).

\bibitem[{\citenamefont{Swindale}(1996)}]{Swindale1996}
\bibinfo{author}{\bibfnamefont{N.~V.} \bibnamefont{Swindale}},
  \bibinfo{journal}{Network: Comput. Neural Syst.}
  \textbf{\bibinfo{volume}{7}}, \bibinfo{pages}{161} (\bibinfo{year}{1996}).

\bibitem[{\citenamefont{Tanaka}(1989)}]{Tanaka1989}
\bibinfo{author}{\bibfnamefont{S.}~\bibnamefont{Tanaka}}, in
  \emph{\bibinfo{booktitle}{Theory of self-organization of cortical maps}},
  edited by \bibinfo{editor}{\bibfnamefont{D.~S.} \bibnamefont{Touretzky}}
  (\bibinfo{publisher}{San Mateo, CA: Morgan Kaufmann}, \bibinfo{year}{1989}),
  pp. \bibinfo{pages}{451--458}.

\bibitem[{Cow(1991)}]{Cowan1991}
\emph{\bibinfo{title}{Simple spin models for the development of ocular
  dominance columns and iso-orientation patches}}, vol.~\bibinfo{volume}{3}
  (\bibinfo{year}{1991}).

\bibitem[{\citenamefont{Cho and Kim}(2004{\natexlab{a}})}]{Cho2004A}
\bibinfo{author}{\bibfnamefont{M.~W.} \bibnamefont{Cho}} \bibnamefont{and}
  \bibinfo{author}{\bibfnamefont{S.}~\bibnamefont{Kim}},
  \bibinfo{journal}{Phys. Rev. Lett.} \textbf{\bibinfo{volume}{92}},
  \bibinfo{pages}{18101} (\bibinfo{year}{2004}{\natexlab{a}}).

\bibitem[{\citenamefont{Cho and Kim}(2004{\natexlab{b}})}]{Cho2004C}
\bibinfo{author}{\bibfnamefont{M.~W.} \bibnamefont{Cho}} \bibnamefont{and}
  \bibinfo{author}{\bibfnamefont{S.}~\bibnamefont{Kim}}
  (\bibinfo{year}{2004}{\natexlab{b}}), \bibinfo{note}{arXiv:q-bio.NC/0405027}.

\bibitem[{\citenamefont{LeVay et~al.}(1985)\citenamefont{LeVay, Connolly,
  Houde, and Essen}}]{LeVay1985}
\bibinfo{author}{\bibfnamefont{S.}~\bibnamefont{LeVay}},
  \bibinfo{author}{\bibfnamefont{D.~H.} \bibnamefont{Connolly}},
  \bibinfo{author}{\bibfnamefont{J.}~\bibnamefont{Houde}}, \bibnamefont{and}
  \bibinfo{author}{\bibfnamefont{D.~C.~V.} \bibnamefont{Essen}},
  \bibinfo{journal}{J. Neurosci.} \textbf{\bibinfo{volume}{5}},
  \bibinfo{pages}{486} (\bibinfo{year}{1985}).

\bibitem[{\citenamefont{Obermayer and Blasdel}(1993)}]{Obermayer1993}
\bibinfo{author}{\bibfnamefont{K.}~\bibnamefont{Obermayer}} \bibnamefont{and}
  \bibinfo{author}{\bibfnamefont{G.~G.} \bibnamefont{Blasdel}},
  \bibinfo{journal}{J. Neurosci.} \textbf{\bibinfo{volume}{13}},
  \bibinfo{pages}{4114} (\bibinfo{year}{1993}).

\bibitem[{\citenamefont{L{\"o}wel and Singer}(1987)}]{Lowel1987a}
\bibinfo{author}{\bibfnamefont{S.}~\bibnamefont{L{\"o}wel}} \bibnamefont{and}
  \bibinfo{author}{\bibfnamefont{W.}~\bibnamefont{Singer}},
  \bibinfo{journal}{Exp. Brain Res.} \textbf{\bibinfo{volume}{68}},
  \bibinfo{pages}{661} (\bibinfo{year}{1987}).

\bibitem[{\citenamefont{Anderson et~al.}(1988)\citenamefont{Anderson,
  Olavarria, and Sluyters}}]{Anderson1988}
\bibinfo{author}{\bibfnamefont{P.~A.} \bibnamefont{Anderson}},
  \bibinfo{author}{\bibfnamefont{J.}~\bibnamefont{Olavarria}},
  \bibnamefont{and} \bibinfo{author}{\bibfnamefont{R.~C.~V.}
  \bibnamefont{Sluyters}}, \bibinfo{journal}{J. Neurosci.}
  \textbf{\bibinfo{volume}{8}}, \bibinfo{pages}{2183} (\bibinfo{year}{1988}).

\bibitem[{\citenamefont{Crowley and Katz}(1999)}]{Crowley1999}
\bibinfo{author}{\bibfnamefont{J.~C.} \bibnamefont{Crowley}} \bibnamefont{and}
  \bibinfo{author}{\bibfnamefont{L.~C.} \bibnamefont{Katz}},
  \bibinfo{journal}{Nature Neurosci.} \textbf{\bibinfo{volume}{2}},
  \bibinfo{pages}{1125} (\bibinfo{year}{1999}).

\bibitem[{\citenamefont{L{\"o}wel et~al.}(1987)\citenamefont{L{\"o}wel,
  Freeman, and Singer}}]{Lowel1987b}
\bibinfo{author}{\bibfnamefont{S.}~\bibnamefont{L{\"o}wel}},
  \bibinfo{author}{\bibfnamefont{B.}~\bibnamefont{Freeman}}, \bibnamefont{and}
  \bibinfo{author}{\bibfnamefont{W.}~\bibnamefont{Singer}},
  \bibinfo{journal}{J. Comp. Neurol.} \textbf{\bibinfo{volume}{255}},
  \bibinfo{pages}{401} (\bibinfo{year}{1987}).

\bibitem[{\citenamefont{Bosking et~al.}(1997)\citenamefont{Bosking, Zhang,
  Schofield, and Fitzpatrick}}]{Bosking1997}
\bibinfo{author}{\bibfnamefont{W.~H.} \bibnamefont{Bosking}},
  \bibinfo{author}{\bibfnamefont{Y.}~\bibnamefont{Zhang}},
  \bibinfo{author}{\bibfnamefont{B.~R.} \bibnamefont{Schofield}},
  \bibnamefont{and}
  \bibinfo{author}{\bibfnamefont{D.}~\bibnamefont{Fitzpatrick}},
  \bibinfo{journal}{J. Neurosci.} \textbf{\bibinfo{volume}{17}},
  \bibinfo{pages}{2112} (\bibinfo{year}{1997}).

\bibitem[{\citenamefont{Grossberg and Olson}(1994)}]{Grossberg1994}
\bibinfo{author}{\bibfnamefont{S.}~\bibnamefont{Grossberg}} \bibnamefont{and}
  \bibinfo{author}{\bibfnamefont{S.~J.} \bibnamefont{Olson}},
  \bibinfo{journal}{Neural Networks} \textbf{\bibinfo{volume}{7}},
  \bibinfo{pages}{883} (\bibinfo{year}{1994}).

\bibitem[{\citenamefont{Takeno and Homma}(1980)}]{Takeno1980}
\bibinfo{author}{\bibfnamefont{S.}~\bibnamefont{Takeno}} \bibnamefont{and}
  \bibinfo{author}{\bibfnamefont{S.}~\bibnamefont{Homma}},
  \bibinfo{journal}{Prog. Theor. Phys.} \textbf{\bibinfo{volume}{64}},
  \bibinfo{pages}{1193} (\bibinfo{year}{1980}).

\bibitem[{\citenamefont{Gouv{\^e}a et~al.}(1989)\citenamefont{Gouv{\^e}a,
  Wysin, Bishop, and Mertens}}]{Gouvea1989}
\bibinfo{author}{\bibfnamefont{M.~E.} \bibnamefont{Gouv{\^e}a}},
  \bibinfo{author}{\bibfnamefont{G.~M.} \bibnamefont{Wysin}},
  \bibinfo{author}{\bibfnamefont{A.~R.} \bibnamefont{Bishop}},
  \bibnamefont{and} \bibinfo{author}{\bibfnamefont{F.~G.}
  \bibnamefont{Mertens}}, \bibinfo{journal}{Phys. Rev. B}
  \textbf{\bibinfo{volume}{39}}, \bibinfo{pages}{11840} (\bibinfo{year}{1989}).

\bibitem[{\citenamefont{Wolf and Geisel}(1998)}]{Wolf1998}
\bibinfo{author}{\bibfnamefont{F.}~\bibnamefont{Wolf}} \bibnamefont{and}
  \bibinfo{author}{\bibfnamefont{T.}~\bibnamefont{Geisel}},
  \bibinfo{journal}{Nature} \textbf{\bibinfo{volume}{395}}, \bibinfo{pages}{73}
  (\bibinfo{year}{1998}).

\bibitem[{\citenamefont{Obermayer et~al.}(1992)\citenamefont{Obermayer,
  Blasdel, and Schulten}}]{Obermayer1992}
\bibinfo{author}{\bibfnamefont{K.}~\bibnamefont{Obermayer}},
  \bibinfo{author}{\bibfnamefont{G.~G.} \bibnamefont{Blasdel}},
  \bibnamefont{and} \bibinfo{author}{\bibfnamefont{K.}~\bibnamefont{Schulten}},
  \bibinfo{journal}{Phys. Rev. A} \textbf{\bibinfo{volume}{45}},
  \bibinfo{pages}{7568} (\bibinfo{year}{1992}).

\bibitem[{\citenamefont{Wolf et~al.}(2000)\citenamefont{Wolf, Pawelzik, Scherf,
  Geisel, and L{\"o}wel}}]{Wolf2000}
\bibinfo{author}{\bibfnamefont{F.}~\bibnamefont{Wolf}},
  \bibinfo{author}{\bibfnamefont{K.}~\bibnamefont{Pawelzik}},
  \bibinfo{author}{\bibfnamefont{O.}~\bibnamefont{Scherf}},
  \bibinfo{author}{\bibfnamefont{T.}~\bibnamefont{Geisel}}, \bibnamefont{and}
  \bibinfo{author}{\bibfnamefont{S.}~\bibnamefont{L{\"o}wel}},
  \bibinfo{journal}{J. Physiol} \textbf{\bibinfo{volume}{94}},
  \bibinfo{pages}{524} (\bibinfo{year}{2000}).

\bibitem[{\citenamefont{L{\"o}wel}(1994)}]{Lowel1994}
\bibinfo{author}{\bibfnamefont{S.}~\bibnamefont{L{\"o}wel}},
  \bibinfo{journal}{J. Neurosci.} \textbf{\bibinfo{volume}{14}},
  \bibinfo{pages}{7451} (\bibinfo{year}{1994}).

\bibitem[{\citenamefont{Albus}(1979)}]{Albus1979}
\bibinfo{author}{\bibfnamefont{K.}~\bibnamefont{Albus}}, \bibinfo{journal}{Exp.
  Brain Res.} \textbf{\bibinfo{volume}{24}}, \bibinfo{pages}{181}
  (\bibinfo{year}{1979}).

\bibitem[{\citenamefont{Diao et~al.}(1990)\citenamefont{Diao, Jia., Swindale,
  and Cynader}}]{Diao1990}
\bibinfo{author}{\bibfnamefont{Y.-C.} \bibnamefont{Diao}},
  \bibinfo{author}{\bibfnamefont{J.}~\bibnamefont{Jia.}},
  \bibinfo{author}{\bibfnamefont{N.~V.} \bibnamefont{Swindale}},
  \bibnamefont{and} \bibinfo{author}{\bibfnamefont{M.~S.}
  \bibnamefont{Cynader}}, \bibinfo{journal}{Exp. Brain Res.}
  \textbf{\bibinfo{volume}{79}}, \bibinfo{pages}{271} (\bibinfo{year}{1990}).

\bibitem[{\citenamefont{Rao et~al.}(1997)\citenamefont{Rao, Toth, and
  Sur}}]{Rao1997}
\bibinfo{author}{\bibfnamefont{S.~C.} \bibnamefont{Rao}},
  \bibinfo{author}{\bibfnamefont{L.~J.} \bibnamefont{Toth}}, \bibnamefont{and}
  \bibinfo{author}{\bibfnamefont{M.}~\bibnamefont{Sur}}, \bibinfo{journal}{J.
  Comp. Neurol.} \textbf{\bibinfo{volume}{387}}, \bibinfo{pages}{358}
  (\bibinfo{year}{1997}).

\bibitem[{\citenamefont{Costa and Costa}(1996)}]{Costa1996}
\bibinfo{author}{\bibfnamefont{J.~E.~R.} \bibnamefont{Costa}} \bibnamefont{and}
  \bibinfo{author}{\bibfnamefont{B.~V.} \bibnamefont{Costa}},
  \bibinfo{journal}{Phy. Rev. B} \textbf{\bibinfo{volume}{54}},
  \bibinfo{pages}{994} (\bibinfo{year}{1996}).

\bibitem[{\citenamefont{Wysin}(1994)}]{Wysin1994}
\bibinfo{author}{\bibfnamefont{W.~M.} \bibnamefont{Wysin}},
  \bibinfo{journal}{Phy. Rev. B} \textbf{\bibinfo{volume}{49}},
  \bibinfo{pages}{8780} (\bibinfo{year}{1994}).

\bibitem[{\citenamefont{Cho and Kim}(2004{\natexlab{c}})}]{Cho2004B}
\bibinfo{author}{\bibfnamefont{M.~W.} \bibnamefont{Cho}} \bibnamefont{and}
  \bibinfo{author}{\bibfnamefont{S.}~\bibnamefont{Kim}}, \bibinfo{journal}{Phy.
  Rev. B} \textbf{\bibinfo{volume}{70}}, \bibinfo{pages}{24405}
  (\bibinfo{year}{2004}{\natexlab{c}}).

\end{thebibliography}

\end{document}